\documentclass[10pt,amsmath,amssymb,aps]{revtex4-2}
\usepackage{graphicx}
\usepackage{bm}
\usepackage{outlines}
\usepackage{natbib}
\usepackage{float}
\usepackage{xcolor}
\usepackage{soul}
\begin{document}

\title{\Large Numerical simulations and universal saturation profiles \\ for viscous fingering patterns in Hele-Shaw flow}
\author{\'Irio M. Coutinho,$^1$\footnote{Present address: Departamento de Engenharia Mecânica, Pontifícia Universidade Católica do Rio de Janeiro, Rio de Janeiro 22451-900, Brazil} Liam C. Morrow,$^2$ and Scott W. McCue$^{3}$}
\email[]{iriomenezes@esp.puc-rio.br, scott.mccue@qut.edu.au}
\affiliation{$^1$Departamento de F\'{\i}sica, Universidade Federal de Pernambuco,
Recife, PE 50670-901, Brazil \\
$^2$Research School of Earth Sciences, Australian National University, Canberra ACT 2601, Australia \\
$^3$School of Mathematical Sciences, Queensland University of Technology, Brisbane QLD 4001, Australia
}

%\date{\today}

\begin{abstract}
Hele-Shaw flows with an interface are known to give rise to complex pattern formation, whereby the Saffman-Taylor instability triggers a viscous fingering process accompanied by tip splitting and branching.  The most popular of these experiments involves a radial configuration with a less viscous fluid injected into a more viscous fluid. In an attempt to characterize the resulting complexity in such an experiment, Beeson-Jones \& Woods~\cite{Beeson-Jones2019} have proposed a type of simple empirical model that aims to predict the saturation profile of the invading fingers as a function of a radial coordinate.  We revisit the proposed saturation model and test its validity over a broad parameter range using fully nonlinear numerical simulations computed with a level set method.  We find that the saturation model is very effective at predicting some near-universal properties of the viscous fingering patterns for one-phase flows, where the invading fluid is neglected, with a sufficiently small surface tension parameter.  For larger values of this parameter and for two-phase flows, there are discrepancies between the model and our observations.  We explain these differences by studying the morphology of the advancing fingers, including pinching at the base and the rate of tip splitting. Overall, our study shows that the Beeson-Jones-Woods saturation model serves as a valid description of DLA-like patterns, but is not universal over two-phase flows, where surface tension and viscosity ratio substantially alter finger morphology and the resulting saturation profile.
\end{abstract}
%\pacs{47.15.gp, 47.54.-r, 47.65.Cb, 47.20.Ma}
\maketitle

\section{Introduction}\label{intro}

Interfacial-flow scenarios in Hele-Shaw cells with two immiscible fluids are extremely well studied, in part because of the visually appealing fingering instabilities, the tip-splitting phenomena and the striking pattern formation \cite{Jaume_Review,LAJEUNESSE2000, Lawless2025NonlinearFingering, Ben-Jacob1990TheGrowth, Langer1989DendritesFormation}.
Other key motivations come from the close links with porous media flow and subsequent applications such as oil recovery \cite{Homsy1987,Saffman1958}.  In the early years, these interfacial patterns were analyzed mostly by experimentalists and, indeed, there is a plethora of such examples \cite{Chen1989,Maher_Review, Tabeling1987AnInstability, Couder1986DendriticExperiment}.
More recently, there are excellent numerical simulations of these phenomena \cite{Fast2006MooresInstability,Li2009}.
%Irio, please include https://doi.org/10.1016/j.jcp.2006.12.023, https://doi.org/10.4208/cicp.OA-2016-0040
To support these experimental and computational results, a challenge for researchers studying these instabilities is to develop simple and reliable measures to characterize the complexity of the fingering patterns. This task is a theme of the present paper.

An example of a Hele-Shaw fingering pattern is provided in Fig.~\ref{fig:introfigure}(a). This figure was produced using the numerical scheme described in the Appendix~\ref{levelset}.  In this simulation, a less viscous fluid is injected into a Hele-Shaw cell that already contains another fluid that is $V=50$ times more viscous (roughly that for water and air), where $V$ is the viscosity ratio between the displaced and displacing fluids [Eq.~(\ref{eq:viscosityratio})]. As is well known, under these circumstances, a circular interface is generally unstable to small perturbations \cite{Paterson1981}, giving rise to the type of fingering pattern that is observed in this image. In general terms, such a Hele-Shaw fingering pattern will be very sensitive to small variations in a number of parameters, most notably
the injection rate of the less viscous fluid, the ratio of fluid viscosities, the initial shape of the bubble, and the surface tension at the interface itself. These effects have been explored at length experimentally and, to a lesser extent, numerically for the standard Hele-Shaw setup that involves injection into a Newtonian fluid between two parallel plates \cite{Fast2006MooresInstability,Li2009} and also for a range of alternative geometries and fluid types \cite{Shelley1997,Zhao2016,Morrow_Numerical,Morrow_Review,Zhao2020,Morrow2023ViscousRotation,Coutinho2025FF}.

\begin{figure}[!ht]
	\centering
 	\includegraphics[width = 1\textwidth]{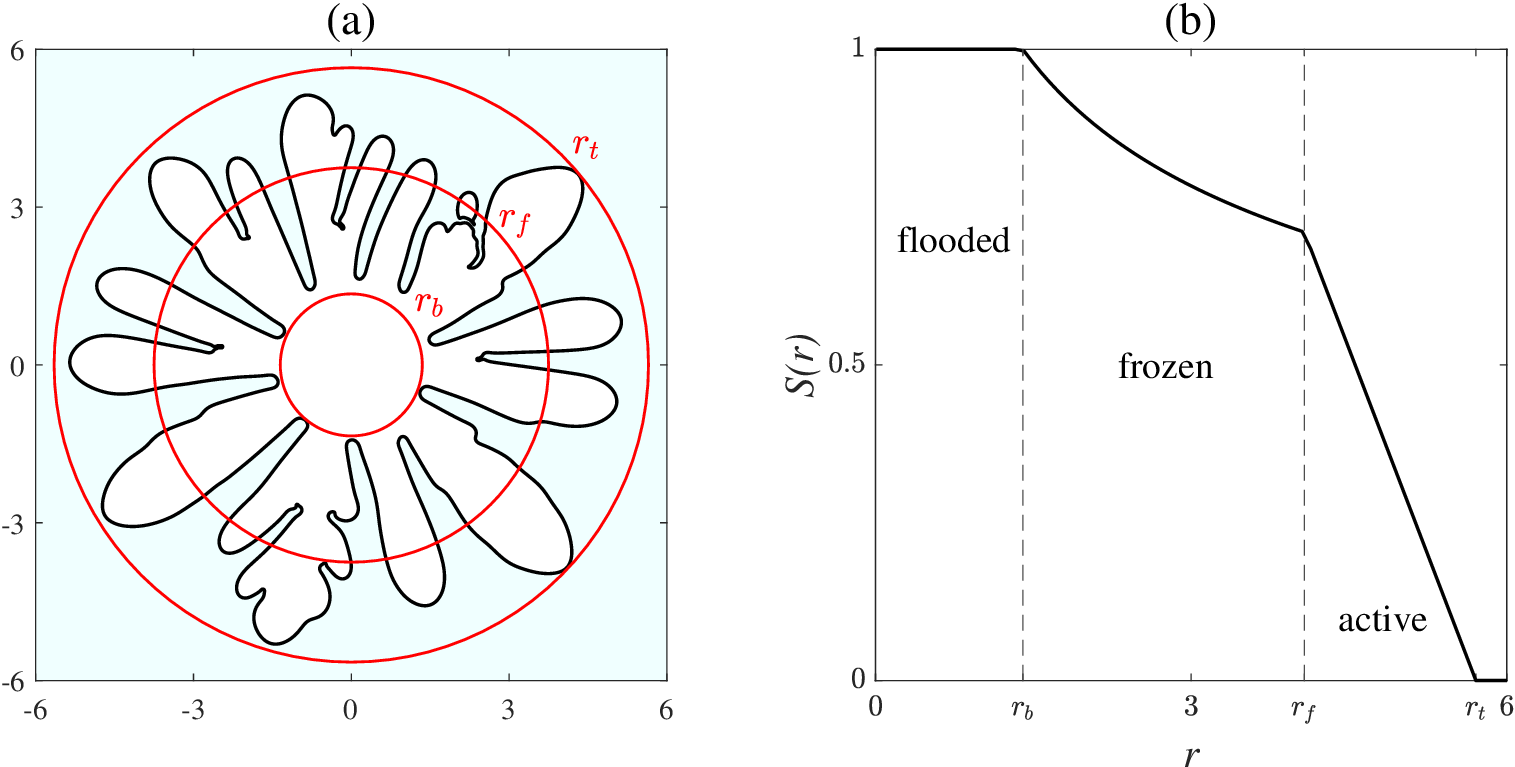}
    	\caption{(a) Numerical simulation of Eqs.~(\ref{laplacep1})-(\ref{source2}) for a viscosity ratio $V=50$ (relevant for air injected into water, where air and water are colored white and blue, respectively) at a fixed dimensionless time $t=50$.  The red circles indicate locations of three important radii:  $r=r_b$, which measures the radius of the innermost fjord and therefore bounds a flooded region filled with the invading fluid only; $r=r_f$, which bounds what Beeson-Jones \& Woods~\cite{Beeson-Jones2019} call the frozen finger zone ($r_b<r<r_f$); and $r_t$, which measures the radius of the outermost tip of the invading fingers and defines what Beeson-Jones \& Woods call the active outer finger zone ($r_f<r<r_t$). (b) A schematic of the saturation profile drawn at $t=50$, indicating the flooded, frozen and active zones.}
	\label{fig:introfigure}
\end{figure}

In this paper, we revisit a simplified model of the viscous fingering pattern in a radial Hele-Shaw cell proposed by Beeson-Jones \& Woods~\cite{Beeson-Jones2019}, referred to as a saturation profile, which characterizes the proportion of a given radius occupied by the invading fluid. For a radial Hele-Shaw experiment of the type illustrated in Fig.~\ref{fig:introfigure}(a), this saturation profile is shown in Ref.~\cite{Beeson-Jones2019} to be well approximated by three regions: an inner region (fully flooded zone) whose radius is constant in time, which is completely occupied by the invading fluid; a second region (frozen finger zone) where the proportion of invading fluid decreases via a fixed power law in radius that is also independent of time; and an outer region (active finger zone) where the proportion decreases linearly to zero.  At a given time, the outer boundary of this outer region is the radius at which the tip of the longest finger is located.  A schematic of such a saturation profile is shown in Fig.~\ref{fig:introfigure}(b).  A key point is that the saturation profile proposed by Beeson-Jones \& Woods provides a relatively straightforward description of complicated viscous fingering patterns. Further, the main conclusion that those authors draw is that there is an apparent universality in this saturation profile, suggesting that despite the significant variability of the interfacial patterns that can arise in a Hele-Shaw experiment, depending on the physical parameters such as injection rate, surface tension and the viscosity ratio, the fundamental measure of radial saturation can be successfully predicted using a very simple formula.

While Beeson-Jones \& Woods have applied their model to radial viscous fingering experiments and diffusion-limited aggregation (DLA) simulations, in this work we assess the universality and validity of the proposed model using fully nonlinear numerical simulations. An advantage of generating numerical solutions is that we are able to cover a much wider parameter space than what is possible in experiments and thereby explore a variety of effects that were not considered by Beeson-Jones \& Woods. As the original Beeson-Jones \& Woods work establishes the validity of their results for highly unstable regimes with viscosity ratios ranging from $V=300$ to $10000$, our study seeks to define the limits of this model by applying it to broader two-phase flow scenarios, including less unstable displacements and systems with comparable fluid viscosities, which were not tested in their original experimental study. Our goals are twofold: (i) to delineate the region of the parameter space in which the empirical saturation model is consistent with our numerical simulations; and (ii) to identify the mechanisms responsible for deviations when they occur. The numerical scheme we use is based on the level set method presented in Morrow et al.~\cite{Morrow_Numerical,Morrow_Review} that was itself based on early work \cite{Hou1997AFlow}. We have extended this one-phase approach to include the effects of both the less and more viscous fluid. Note that most of the other numerical methods for simulating fully nonlinear one- and two-phase Hele-Shaw flows involve either a boundary-integral formulation \cite{Dai1993,Hou2001BoundaryMaterials,Fast2006MooresInstability,Li2009,Zhao2020} or a finite-element method~\cite{Vaquero-Stainer2019Self-similarDepth, Pihler-Puzovic2013ModellingCells}, which differs from our presented level set scheme.

In section~\ref{equations}, we present the governing equations for two-phase Hele-Shaw flow, and show that the setup is characterized by two dimensionless parameters, namely the effective surface tension $\sigma$ and the viscosity ratio $V$. Our paper continues in section~\ref{discussion}, where we solve the governing equations numerically to assess the validity of the Beeson-Jones-Woods model across a broad parameter range, for both one- and two-phase flows. The details of the numerical scheme are presented in the Appendix~\ref{levelset}. Finally, we end in section~\ref{conclude} with a brief discussion summarizing our findings.

\section{Governing Equations}\label{sec:twophase}
\label{equations}
\begin{figure}[!ht]
	\centering
 	\includegraphics[width = 0.4\textwidth]{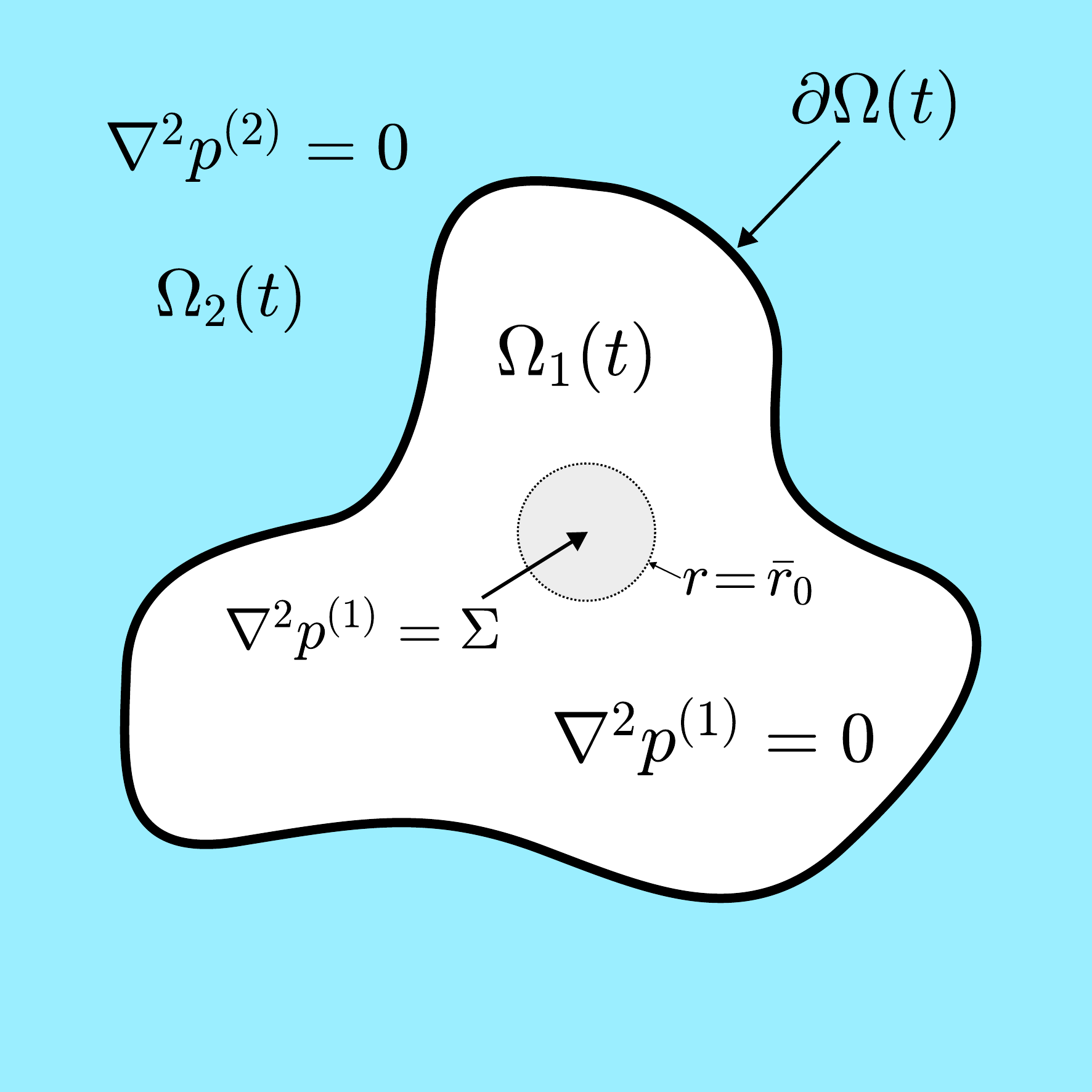}
        \caption{Schematic representation of the two-phase radial Hele-Shaw flow. A central droplet (fluid 1) with viscosity $\mu_1$ occupies the inner region $\Omega_1(t)$ and displaces a more viscous outer fluid (fluid 2, $\mu_2 > \mu_1$) in region $\Omega_2(t)$. The fluids are separated by a sharp interface $\partial \Omega(t)$. Fluid 1 is injected at a constant flow rate $Q$ from a central source $\Sigma$ with a smoothing radius $\bar{r}_0$.}
	\label{fig:schematic}
\end{figure}

We consider a Hele-Shaw cell of gap spacing, $b$, containing two immiscible, Newtonian and incompressible fluids. In this radial configuration, a viscous fluid (fluid 2) is displaced by another of lesser viscosity (fluid 1), which is injected through a small circular region at a constant flow rate $Q$. The viscosities are given by $\mu_k$, where $k=1,2$, and $\mu_2 > \mu_1$. The two fluids are separated by a sharp interface with surface tension $\gamma$. A schematic representation of the model is presented in Fig.~\ref{fig:schematic}.

Under these conditions, the flow is governed by Darcy's law~\cite{Homsy1987, Maher_Review, Jaume_Review}
\begin{equation}
\label{darcy1}
    \mathbf{v}_k = - \frac{b^2}{12 \mu_k} \nabla p^{(k)}, \; \mathbf{x} \in \Omega_k(t).
\end{equation}
where $\mathbf{v}_k$, $p^{(k)}$ and $\Omega_k(t)$ denote the gap-averaged velocity, the pressure fields, and the connected domain of fluid $k = 1,2$. Note that $\Omega_2 = \mathbb{R}^2 \setminus \Omega_1$. The interface between the two fluids is denoted by $\partial \Omega(t)$. Taking the divergence of Darcy's law and enforcing the incompressibility condition ($\nabla \cdot \mathbf{v}_k = 0$), we see that the pressure field obeys Poisson's equation
\begin{equation}
\label{laplace1}
    \nabla^2 p^{(k)} = \frac{12 \mu_k}{b^3}\Sigma, \; \mathbf{x}\in \Omega_k(t),
\end{equation}
where
\begin{equation}
\label{source1}
   \Sigma = \begin{cases}
    \displaystyle\frac{Q}{b \bar{r}^2_0} \left(1 + \cos \left(\frac{\pi r}{\bar{r}_0}\right)\right), \; \text{if } r\leq \bar{r}_0,\\
    0 , \; \text{if } r > \bar{r}_0 .\end{cases}
\end{equation}
acts as a source term around the origin. Equation~(\ref{source1}) was proposed by Hou \textit{et. al}~\cite{Hou1997AFlow}, and arises from a smoothed Dirac delta function, where $\bar{r}_0$ represents a smoothing radius. Furthermore, a far-field boundary condition
\begin{equation}
\label{farfield1}
    \frac{b^3}{12 \mu_2} \frac{\partial p^{(2)}}{\partial r} \sim -\frac{Q}{2\pi r}, \; r \to \infty,
\end{equation}
where $r = |\mathbf{x}|$, acts as a sink at infinity~\cite{Morrow_Review,Morrow_Numerical}.

To fully specify our problem, two boundary conditions on the interface are required. The first is the kinematic boundary condition, which states that the normal component of the velocity is continuous across the interface~\cite{Homsy1987, Maher_Review, Jaume_Review},
\begin{equation}
\label{kinematic1}
    \mathbf{v}_1 \cdot \mathbf{n} = \mathbf{v}_2 \cdot \mathbf{n}, \; \mathbf{x} \in \partial \Omega(t),
\end{equation}
where $\mathbf{n}$ is the unit normal vector to the interface. This condition can be rewritten as
\begin{equation}
\label{kinematic}
    \mathbf{v}_k \cdot \mathbf{n} = - \frac{b^2}{12 \mu_k} \frac{\partial p_k}{\partial n}, \; \mathbf{x} \in \partial \Omega(t),
\end{equation}
relating the normal velocity at the interface with the normal pressure gradient. Furthermore, due to surface tension effects, the pressure field is discontinuous across the interface, giving rise to a dynamical boundary condition described by the Young-Laplace's equation~\cite{Homsy1987, Maher_Review, Jaume_Review},
\begin{equation}
\label{younglaplace1}
    p^{(2)} - p^{(1)} = - \gamma \kappa, \; \mathbf{x} \in \partial \Omega(t),
\end{equation}
where $\kappa$ is the in-plane curvature of the interface and $\gamma$ the surface tension.

We nondimensionalize Eqs.~(\ref{laplace1})-(\ref{younglaplace1}) by scaling lengths by the average initial radius of the interface $r_0$ and time by $T = r_0^2 b /Q$, which yields
\begin{equation}
\label{laplacep1}
    \nabla^2 p^{(1)} = \frac{\Sigma}{V}, \; \mathbf{x} \in \Omega_1(t),
\end{equation}
\begin{equation}
\label{laplacep2}
    \nabla^2 p^{(2)} = 0, \; \mathbf{x} \in \Omega_2(t),
\end{equation}
\begin{equation}
\label{younglaplace2}
    p^{(2)} - p^{(1)} = - \sigma \kappa, \; \mathbf{x} \in \partial\Omega(t),
\end{equation}
\begin{equation}
\label{kinematic2}
    \mathbf{v}_1 \cdot \mathbf{n} = \mathbf{v}_2 \cdot \mathbf{n}, \; \mathbf{x} \in \partial \Omega(t),
\end{equation}
\begin{equation}
\label{farfield2}
    \frac{\partial p^{(2)}}{\partial r} \sim -\frac{1}{2\pi r}, \; r \to \infty,
\end{equation}
where
\begin{equation}
    \sigma = \frac{b^3 \gamma}{12 \mu_2 Q r_0^2}
\label{eq:surfacetension}
\end{equation}
is an effective surface tension,
\begin{equation}
V = \frac{\mu_2}{\mu_1}
\label{eq:viscosityratio}
\end{equation}
is the viscosity ratio, and
\begin{equation}
\label{source2}
    \Sigma = \begin{cases}
    \displaystyle\frac{1}{\bar{r}^2_0} \left(1 + \cos \left(\frac{\pi r}{\bar{r}_0} \right)\right), \; \text{if } r\leq \bar{r}_0,\\
    0 , \; \text{if } r > \bar{r}_0.\end{cases}
\end{equation}
For the entirety of this work, we choose $\bar{r}_0 = 0.05$, which ensures that $\Sigma$ is always zero in $\Omega_2(t)$. To solve Eqs.~(\ref{laplacep1})-(\ref{source2}), we employ a two-phase, sharp interface, level set scheme, described in the Appendix~\ref{levelset}. We consider that initially ($t = 0$), the fluid-fluid interface is situated at~\cite{Cardoso1995,Morrow_Numerical, Morrow_Review, Coutinho2025rot, Coutinho2026SuppressingSimulations}
\begin{equation}
\label{inital}
    \mathcal{R}(\theta,0) = 1 + 10^{-2} \times \sum_{n=2}^{20} \cos[ n(\theta - \varphi_n)] ,
\end{equation}
where $\varphi_n$ is a random phase ranging between 0 and $2\pi$, that mimics the presence of a constant low level of random noise in the system, which may arise due to inhomogeneities of the Hele-Shaw plates, or thermal and pressure fluctuations~\cite{Cardoso1995}.

The set of dimensionless equations~(\ref{laplacep1})-(\ref{source2}) fully describes our two-phase system, which is characterized by two dimensionless parameters. The first parameter is the effective surface tension $\sigma$, which quantifies the relative importance of surface tension compared with viscous effects. Typically, larger values of $\sigma$~[Eq.~(\ref{eq:surfacetension})] lead to a less unstable interface~\cite{Homsy1987, Paterson1981, Maher_Review, Jaume_Review}. This parameter can vary over several orders of magnitude, and can be easily tuned experimentally for a fixed pair of fluids by adjusting the imposed flow rate $Q$, the initial droplet radius $r_0$, or the gap thickness $b$. The second parameter is the viscosity contrast $V$~[Eq.~(\ref{eq:viscosityratio})], which measures the relative viscosity of the two fluids. In the one-phase limit, where an inviscid fluid displaces a viscous one, $V \to \infty$. We stress that the physical parameters used in this work are consistent with those commonly employed in experimental radial Hele-Shaw cell investigations~\cite{Homsy1987, Paterson1981, Maher_Review, Jaume_Review}. Although we attempt to match the parameters considered by Beeson-Jones \& Woods~\cite{Beeson-Jones2019}, their work does not report all dimensional quantities required to compute the effective surface tension $\sigma$, most notably the initial drop radius $r_0$, whose square appears in the denominator of $\sigma$ (see Eq. (\ref{eq:surfacetension})).

\section{Analysis of the empirical saturation model}
\label{discussion}

\subsection{Summary of Beeson-Jones-Woods model}

Aiming to succinctly illustrate the radial viscous fingering instabilities in a Hele-Shaw cell, Beeson-Jones \& Woods~\cite{Beeson-Jones2019} present an empirical saturation profile model for Hele-Shaw flow, describing the behavior of the azimuthally-averaged fraction of the area occupied by the invading fluid, $S(r,t)$. The saturation $S(r,t)$ is computed through the area covered by the displacing fluid in an annular ring delimited by circles of radii $r$ and $r + \delta r$, $2 \pi r S(r,t) \delta r$, with $\delta r \ll r$. As discussed in Sec.~\ref{intro}, their model divides the saturation profile into three zones, illustrated in Fig.~\ref{fig:introfigure}. For $r < r_b$, we have a central \textit{fully flooded region}, where the injected fluid occupies the entire region, leading to $S = 1$. The radius $r_b$ delimits this flooded region and, while dependent on the physical parameters of the fluids and cell, is assumed by Beeson-Jones \& Woods to be constant in time. For $r_b \leq r < r_f(t)$, we have a \textit{frozen finger zone}, where the saturation $S(r,t)$ is postulated to be independent of time and decays in a power-law fashion, with exponent $D - 2$, where $D$ is the fractal dimension of the pattern. Finally, we have an \textit{active finger zone} for $r_f(t) \leq r < r_t(t)$, in which the fingers advance in time, thicken, and bifurcate. In this zone, some data suggests that the saturation decays linearly with $r$. The radius $r = r_t(t)$ represents the maximum radius of the pattern, thus $S(r,t) = 0$ for $r \geq r_t(t)$.

Mathematically, the Beeson-Jones-Woods model is expressed as~\cite{Beeson-Jones_Thesis}
\begin{equation}
\label{model}
    S(r,t)_\textrm{model} =
    \begin{cases}
        1, & \text{for } 0 < r < r_b, \; (\textrm{flooded region}) \\[0.5em]
        \left( \dfrac{r}{r_b} \right)^{D-2}, & \text{for } r_b \leq r < r_{f}(t),  \; (\textrm{frozen region}) \\[0.5em]
        \left( \dfrac{r_f}{r_b} \right)^{D-2} \dfrac{\Lambda - r/r_f(t)}{\Lambda - 1}, & \text{for } r_f(t) \leq r < r_t(t),  \; (\textrm{active region}) \\[0.5em]
        0, &
        \text{otherwise,}
    \end{cases}
\end{equation}
where, $\Lambda \equiv r_t(t) / r_f(t)$ and $r_b$ are fitting parameters. For the viscous fingering pattern to represent a fractal with constant fractal dimension, it is expected that $\Lambda$ and $r_b$ must be constants~\cite{Beeson-Jones2019}.

Employing this formulation, Beeson-Jones \& Woods compare the model (\ref{model}) to data from injection-driven radial Hele-Shaw flow experiments and from DLA simulations. For the DLA simulations, considering $10^6$ off-lattice walkers, they report that the best fitting parameters are $\Lambda = 1.43 \pm 0.04$ and $r_b = 0.14 \pm 0.01$, when the fractal dimension is fixed at $D = 1.7$. For the several experiments in radial Hele-Shaw flow, both new and from the literature, across viscosity ratios ranging from $V = 300$ to $2 \times 10^4$, where $V$ is defined in Eq.~(\ref{eq:viscosityratio}), they consistently found $\Lambda$ values in the range $1.43$--$1.46$. Notably, they have not provided any measure of the fractal dimension of the patterns resulting from the performed experiments, but instead simply set $D = 1.7$ for every case. While this value is commonly accepted as the fractal dimension of both viscous fingering and DLA patterns, in the case of viscous fingering other values of $D$ have also been reported, ranging from $1.64$--$1.85$~\cite{May1989FractalPatterns, Couder1988ViscousGeometry, Mly1987DynamicsMedia, Rauseo1987DevelopmentPatterns, Praud2005}. The optimal fit is obtained by minimizing the error
\begin{equation}
\label{error}
E = \frac{\int  2\pi r |S_\textrm{model} - S| \textrm{d}r}{\int 2\pi r S \textrm{d}r},
\end{equation}
where $S$ refers to the saturation obtained from the experimental or numerical data, and $S_\textrm{model}$ to Eq.~(\ref{model}).

In their work, Beeson-Jones \& Woods present evidence for a universal saturation profile for radial Hele-Shaw flows, which allows one to concisely describe the evolution of viscous fingering. Here, our goal is to apply our numerical scheme to evaluate the universality and validity of this model. A key advantage of the numerical approach is that it allows the exploration of a much broader parameter space than is accessible experimentally, including lower viscosity ratios.

\subsection{One-phase flows}

\begin{figure}[!ht]
	\centering
	\includegraphics[width = \textwidth]{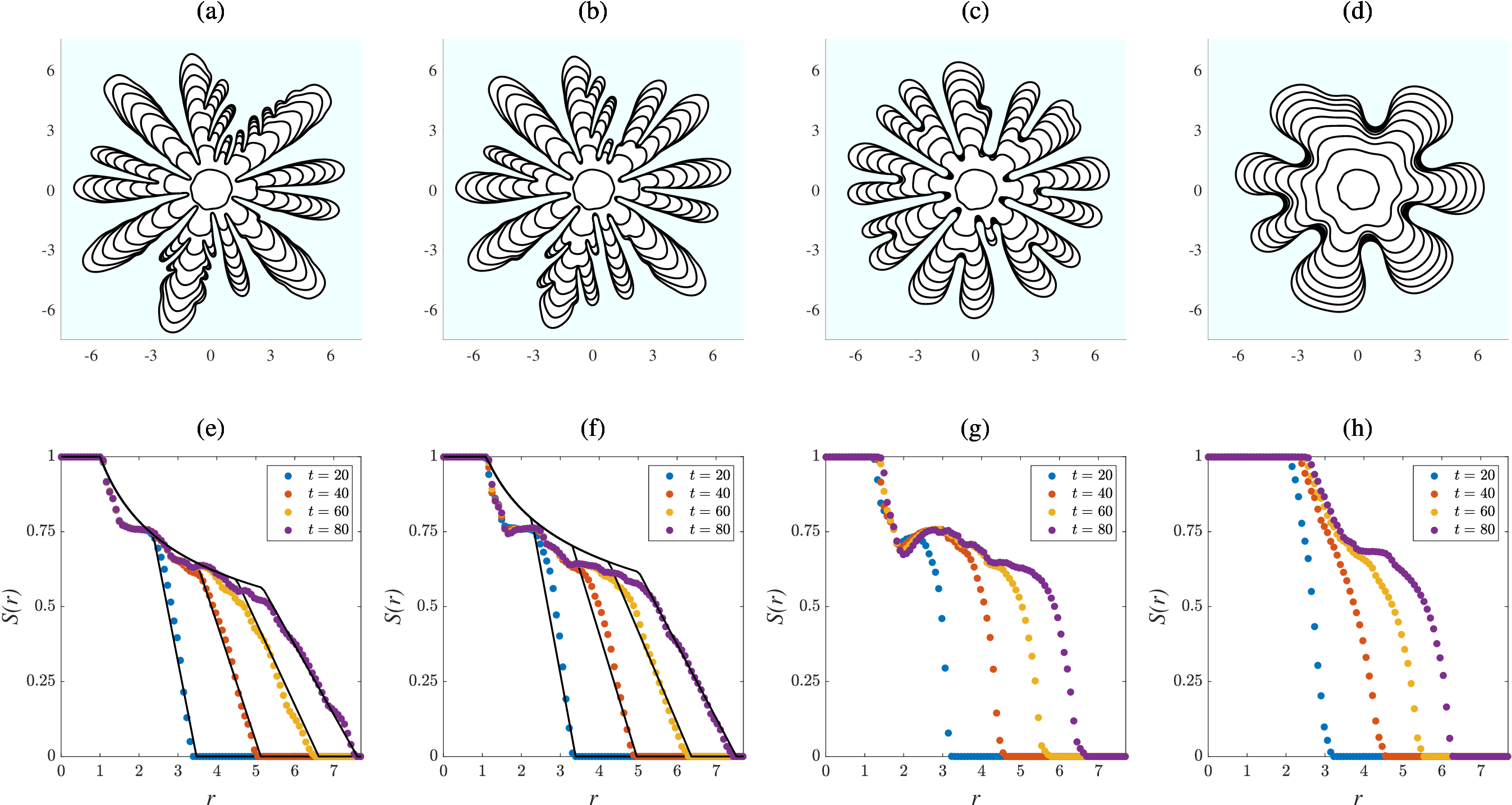}
	\caption{(a)-(d) Time evolution of the interfacial patterns obtained numerically with the level set scheme, for one-phase flow ($V = \infty$). The patterns are presented for $0 \leq t \leq 80$.  (e)-(h) Corresponding saturation profiles of the patterns presented. For (e) and (f), a fit with the Beeson-Jones-Woods model is graphed in black. The simulations are performed for (a) and (e): $\sigma = 2 \times 10^{-6}$, (b) and (f): $\sigma = 2 \times 10^{-5}$, (c) and (g): $\sigma = 2 \times 10^{-4}$, and (d) and (h): $\sigma = 2 \times 10^{-3}$. The measured fractal dimension $D$ of each pattern is (a) $D = 1.65$, (b) $D = 1.68$, (c) $D = 1.79$, and (d) $D = 1.99$.}
	\label{onephase}
\end{figure}

In this section, we explore one-phase flows, presented in Fig.~\ref{onephase}, where a viscous fluid is displaced by an inviscid one ($V = \infty$). This configuration is appropriate to begin our investigation of the Beeson-Jones-Woods model, as the experiments analyzed in Ref.~\cite{Beeson-Jones2019} were conducted at high viscosity ratios, with $V=300$ or higher. The first row of Fig.~\ref{onephase} [(a)-(d)] presents interfacial patterns generated from our numerical scheme, for four values of the effective surface tension:  (a) $\sigma = 2 \times 10^{-6}$, (b) $\sigma = 2 \times 10^{-5}$, (c) $\sigma = 2 \times 10^{-4}$, and (d) $\sigma = 2 \times 10^{-3}$. Panels (a)-(d) show the full temporal evolution of the interface $0 \leq t \leq t_f$, where $t_f = 80$. The patterns are presented in equal time spacings $\delta t = 10$. The second row of Fig.~\ref{onephase} [(e)-(h)] reports the corresponding saturation profiles, $S(r,t)$, plotted as functions of $r$ at $t = 20$, 40, 60, and 80. In Figs.~\ref{onephase}(e) and (f), the solid black line represents the fit to the model given in Eq.~(\ref{model}). For the level set formulation, the quantity $S(r,t)$ is calculated as
\begin{equation}
    S(r,t) = \frac{1}{2 \pi r \delta r}\int_{\psi(r)} H[\phi(\mathbf{x})] d\mathbf{x},
\end{equation}
where $\psi(r)$ is the annular region delimited by circles of radii $r$ and $r + \delta r$, and $H[\phi(\mathbf{x})]$ is a smeared-out Heaviside function~\cite{OsherBook}

% Requires: \usepackage{amsmath}

\begin{equation}
    H(\phi) =
    \begin{cases}
        0, & \phi < -\epsilon, \\[4pt]
        \dfrac{1}{2} + \dfrac{\phi}{2\epsilon} + \dfrac{1}{2\pi} \sin\!\left( \dfrac{\pi \phi}{\epsilon} \right), & -\epsilon \le \phi \le \epsilon, \\[4pt]
        1, & \epsilon < \phi,
    \end{cases}
    \label{eq:placeholder_label}
\end{equation}
where $\epsilon = 1.5 \Delta r$, where $\Delta r$ is the grid spacing in the $r$ direction. When calculating $S(r,t)$ from our numerical data, we have detected no significant change in our results for different values of $\delta r$, and set it to $\delta r = 5 \Delta r$ for the remainder of this work (reference~\cite{Beeson-Jones2019} makes no mention of their choice of $\delta r$ for their experimental measures).

Figure~\ref{onephase}(a) shows the time evolution of the growing interface, computed using the level set scheme, for the lowest effective surface tension considered, $\sigma = 2 \times 10^{-6}$. At this low value of surface tension, the initially growing fingers rapidly develop stationary fjords (also called finger bases), delineating a fully flooded region of fixed radius $r_b$. The fingers continue to advance and undergo repeated side-branching, where small fingers branch out to the side of a main finger. An interfacial pattern is produced with the characteristic features of the Saffman-Taylor instability, including several tip-splitting events and strong variability in finger length. These effects lead to a highly branched pattern. Following May \& Maher~\cite{May1989FractalPatterns}, we compute the fractal dimension $D$ by comparing the logarithm of the radius of gyration $R_g$ with the logarithm of the pattern area, using the expected scaling $R_g \sim A^{1/D}$. This procedure yields $D = 1.65$ for the pattern shown in Fig.~\ref{onephase}(a).

The saturation profile of this fingering pattern is presented in Fig.~\ref{onephase}(e). Solid circles denote the numerical data and the black line represents the corresponding model fit. For this small value of surface tension, the saturation profile closely follows the Beeson-Jones-Woods model. At small radii, a fully saturated flooded region is observed, where $S(r,t) = 1$. Note that this region is delimited by a radius $r_b$ constant in time, corresponding to the finger bases being stationary. The saturation then transitions into a frozen region, where the saturation is time-independent and decays with $r$, according to Eq.~(\ref{model}). While some oscillations appear on the curve, the overall behavior is monotonically decreasing. Finally, an active zone appears, characterized by a linear dependence of the saturation on the radius [Eq.~(\ref{model})].

To fit the saturation profile data to the model, we incorporate the measured fractal dimension $D$, rather than fixing it at $D = 1.7$ as was done in Ref.~\cite{Beeson-Jones2019}. We obtain the best fit by minimizing the error quantity defined in Eq.~(\ref{error}) at $t = t_f$. For the pattern presented in Fig.~\ref{onephase}(a), at $\sigma = 2\times 10^{-6}$ with $D = 1.65$, the fit yields $\Lambda = 1.47$ and $r_b = 1$, with an error of approximately 4\% [$E = 0.040$]. This value of $\Lambda$ is close to the range $1.43$-$1.46$ reported by Beeson-Jones and Woods, indicating an agreement between our numerical simulations and the Beeson-Jones-Woods model, for this set of parameters.

Increasing the effective surface tension to $\sigma = 2\times 10^{-5}$, as shown in Figs.~\ref{onephase}(b) and (f), yields results qualitatively similar to those in Figs.~\ref{onephase}(a) and (e). The interfacial pattern exhibits multiple finger tip bifurcations and remains highly branched, still presenting a few side-branching events. The measured fractal dimension was $D = 1.68$. The corresponding saturation profile again shows well-defined flooded, frozen and active zones. Fitting this profile to the model gives $\Lambda = 1.50$, $r_b = 1.1$, and $E = 0.056$, values that remain reasonably close to those reported in Ref.~\cite{Beeson-Jones2019}.

In contrast, further increasing the effective surface tension to $\sigma = 2 \times 10^{-4}$ [Figs.~\ref{onephase}(c) and (g)] produces a visually distinct fingering pattern, where side-branching events are suppressed. Even though many finger tip-splitting events still occur, the fractal dimension rises substantially to $D = 1.79$; a closer inspection also reveals that the tips of the fjords do not remain completely stationary, but instead advance slowly over time, which is in stark contrast with the Beeson-Jones \& Woods model, violating their assumption that $r_b$ is constant. While the saturation profile continues to display features resembling flooded, frozen and active zones, its overall behavior no longer follows the empirical model. In particular, immediately after the flooded region, the saturation exhibits an abrupt drop followed by a rise, creating a local minimum that was not present at lower values of $\sigma$ in Figs.~\ref{onephase}(e) and (f). Given the markedly different curve shape of $S(r,t)$, no satisfying fit was found for this fingering pattern.

Finally, for the highest effective surface tension considered, $\sigma = 2 \times 10^{-3}$ [Figs.~\ref{onephase}(d) and (h)], the system develops a distinct interfacial pattern characterized by short, broad viscous fingers. Finger branching is almost entirely absent, with only a few tip-splitting events appearing, yielding a fractal dimension of $D = 1.99$, and the saturation profile deviating from the empirical model. At early times, the saturation abruptly goes from fully saturated to zero, whereas at later times a short plateau emerges in which the saturation remains nearly constant with radius. Under this set of parameters, there is no indication of a frozen zone forming.

The numerical simulations in Fig.~\ref{onephase} indicate that, for one-phase flows, the model proposed by Beeson-Jones \& Woods is only valid for patterns formed at low surface tension values. To begin to understand this restriction, it is worth reflecting on the original motivation for the saturation model, which appears to come from comparisons with DLA (see Beeson-Jones' PhD thesis~\cite{Beeson-Jones_Thesis}). In DLA, particles undergoing random motion stick to a growing structure upon contact, forming a cluster whose boundary is the set of outermost sites exposed to the surrounding medium, accessible to incoming particles. The probability distribution of a random walker attaching to this boundary is determined by solutions of Laplace’s equation. This corresponds to the likelihood that diffusion reaches each exposed location and matches the flux distribution associated with Hele-Shaw flow. Thus, DLA can serve as a model that reproduces key features of Hele-Shaw flow. However, this correspondence holds only when our surface tension parameter is small, with values below $\sigma = 10^{-5}$~\cite{May1989FractalPatterns, Couder1988ViscousGeometry, Mly1987DynamicsMedia, Rauseo1987DevelopmentPatterns, Praud2005}. At higher surface tension parameter values, the resulting viscous fingering patterns depart from DLA-like behavior, and the saturation profiles accordingly deviate from the empirical model. A few works have tried to mimic surface tension in DLA simulations by changing the attaching probability of the walkers~\cite{King1987TheMedia, Zhang1998StudyModel}, but this case is not considered either by Ref.~\cite{Beeson-Jones2019} or in this work.

\begin{figure}[!ht]
	\centering
	\includegraphics[width = \textwidth]{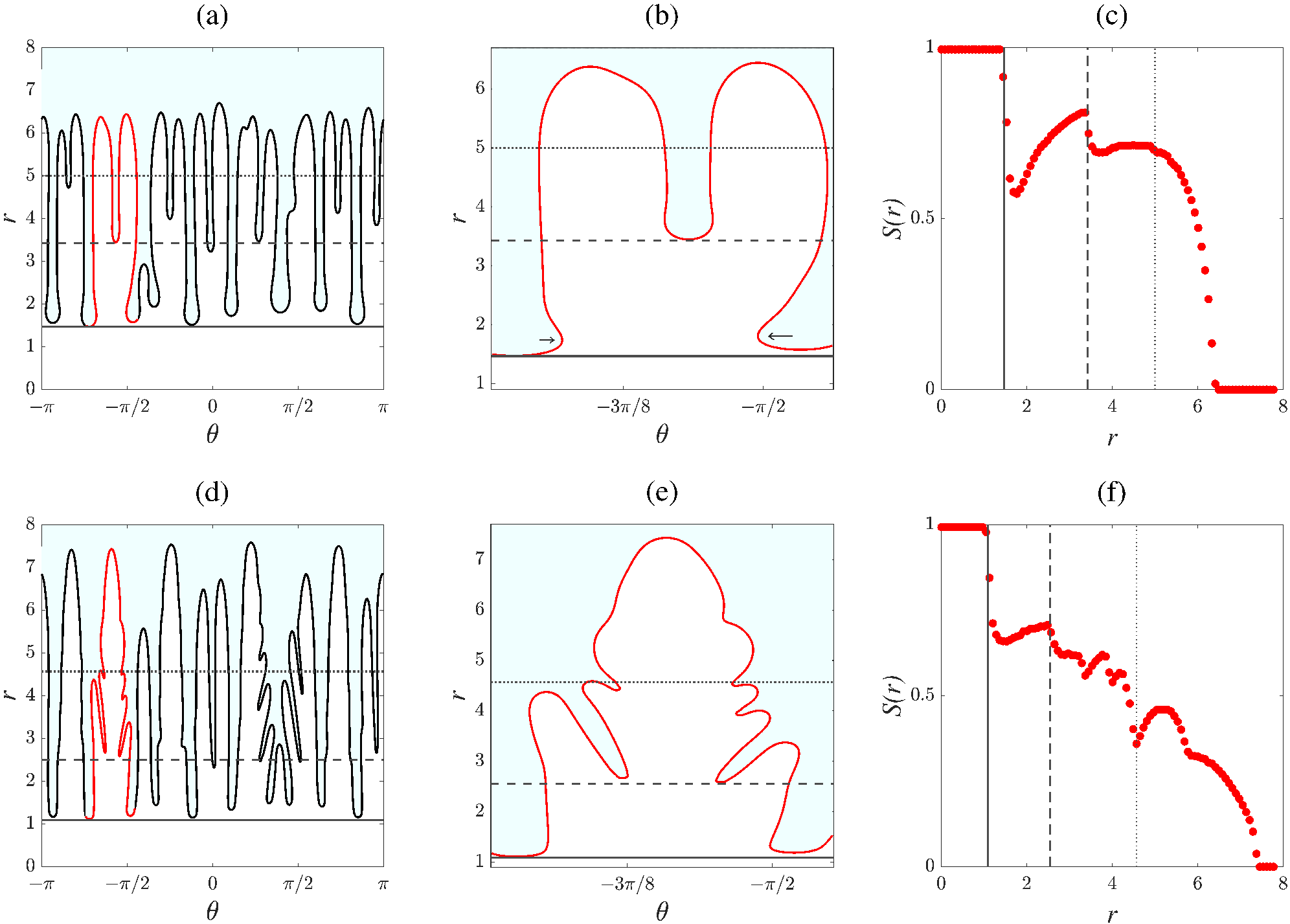}
    	\caption{(a) and (d): Polar representation of a snapshots of the interfaces portrayed in Figs~\ref{onephase}(c) and (a) at $t = 80$, respectively. A close-up of a selected viscous finger, highlighted in red, is presented in (b) and (e). (c) and (d) presents the normalized saturation profile of the highlighted fingers. Small black arrows on (b) indicate a ``clamping" effect on the viscous finger.}
	\label{open}
\end{figure}

\subsection{Effects of finger morphology}

In this section, we investigate why the Beeson-Jones \& Woods model is unable to describe the observed saturation profiles as $\sigma$ increases. In Figs.~\ref{open}(a) and (d), we plot the interface radius as a function of the polar angle $\theta$ for the final interfacial patterns presented in Figs.~\ref{onephase}(c) [$\sigma = 2\times 10^{-4}$] and (a) [$\sigma = 2\times 10^{-6}$], respectively, at $t = t_f = 80$. Figures~\ref{open}(b) and (e) represent a close-up of a single finger of Figs.~\ref{open}(a) and (d), indicated by the red contour. Finally, Figs.~\ref{open}(c) and (f) represent the normalized saturation profile for the single finger presented in Figs.~\ref{open}(b) and (e). Throughout Fig.~\ref{open}, three thin, black lines (solid, dashed and dotted) guide the eye to relevant events.

We begin our discussion with the first row of Fig.~\ref{open}, corresponding to $\sigma = 2 \times 10^{-4}$ [Figs.~\ref{open}(a)-(c)]. While inspection of the full pattern in Fig.~\ref{open}(a) shows several tip-splitting events, most of those occur on the radial interval $3 \lesssim r \lesssim 5$. Finger lengths exhibit little variability, with most terminating near $r \approx 6$. The impact of the finger morphology on the saturation profile can be understood by examining the close-up of the finger highlighted in red in Fig.~\ref{open}(a), shown in Fig.~\ref{open}(b), together with its corresponding saturation profile in Fig.~\ref{open}(c). Initially, the finger displays a flooded zone from $r = 0$ to approximately $r \approx 1.4$, as indicated by the solid line. Beyond this point, between the solid and dashed lines, $S(r,t)$ drops sharply before undergoing a significant rise in its value. As seen in Fig.~\ref{open}(b), this behavior reflects a pronounced “clamping” of the finger base, where the finger quickly narrows and widens, indicated by small black arrows. A second abrupt drop in saturation occurs around $r \approx 3.4$, coinciding with a tip-splitting event (dashed line). Afterward, the saturation forms a short plateau and then transitions into a steady decline (dotted line) as $r$ reaches the finger tip.

In the second row of Fig.~\ref{open}, for $\sigma = 2 \times 10^{-6}$ [Figs.~\ref{open}(d)-(f)], we see from Fig.~\ref{open}(d) that branching occurs along the entire extent of the fingers, unlike the higher-surface tension case. The fingers also present more length variability. The close-up of the finger marked in red, shown in Fig.~\ref{open}(e), reveals that the “clamping” effect is much more subtle. In Fig.~\ref{open}(f), the decrease in saturation in the post flooded zone, indicated by the solid line, is present but far less pronounced than in Fig.~\ref{open}(c). Likewise, the sharp rise observed at lower surface tension becomes far more subtle at higher surface tension. Due to multiple side-branching events occurring along the same finger, the resulting saturation curve exhibits significant oscillations, preventing the appearance of clear plateaus or abrupt transitions.

From Fig.~\ref{open}, we can begin to understand the morphological reasons why the Beeson-Jones-Woods model appears to fail at higher surface tension parameters. At low surface tension, finger bases are relatively smooth, whereas at higher surface tension they exhibit a pronounced “clamping” near the base, which produces the local minimum seen in Fig.~\ref{onephase}(g). Figures~\ref{open}(b) and (c) shows that tip-splitting events generate abrupt changes in the saturation profile. For low surface tension, however, side-branching is frequent and distributed along the entire extent of the fingers, and, supplemented by tip-splitting events, causes these abrupt variations to be effectively averaged out. In contrast, at higher surface tension, side-branching is not present, and bifurcations cluster within a limited radial interval. As a result, for higher surface tensions the saturation profile may show sharper transitions and a plateau, such as the one in Fig.~\ref{onephase}(h).

\begin{figure}[!ht]
	\centering
	\includegraphics[width = \textwidth]{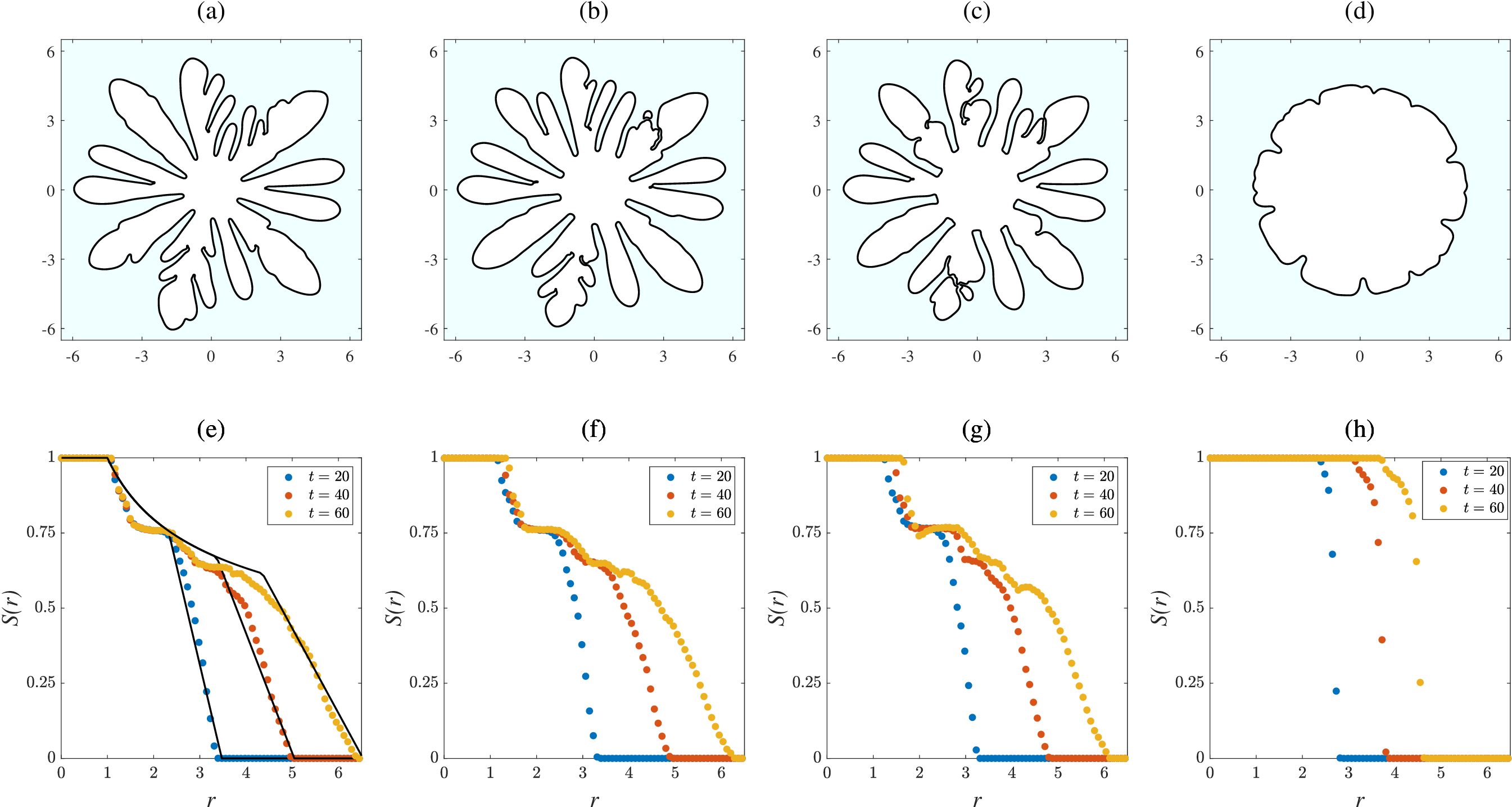}
    	\caption{(a)-(d) Snapshots of the two-phase interfacial patterns obtained numerically with the level set scheme, for $\sigma = 2 \times10^{-6}$, at $t = 60$. (e)-(h) Corresponding saturation profiles of the patterns presented. The simulations are performed for the following viscosity contrasts: (a) and (e): $V=300$, (b) and (f): $V=50$, (c) and (g): $V=25$, and (d) and (h): $V=2$.}
	\label{twophase}
\end{figure}

\subsection{Two-phase flows}

Having established the main features of the saturation profile for one-phase flow, we turn our attention to two-phase flow. In Fig.~\ref{twophase}, we consider four values of the viscosity ratio: $V = 300$ for \ref{twophase}(a) and (e), $V = 50$ for \ref{twophase}(b) and (f), $V = 25$ for \ref{twophase}(c) and (g), and $V = 2$ for \ref{twophase}(d) and (h). The simulations are performed for the representative value of the effective surface tension $\sigma = 2 \times 10^{-6}$, which was chosen since the corresponding result for one-phase flow better matches the Beeson-Jones-Woods model. Snapshots of the interfacial patterns at $t = t_f = 60$ are presented in the first row of Fig.~\ref{twophase}, while the second row contains the corresponding saturation curves.

The first viscous fingering pattern in Fig.~\ref{twophase}(a), corresponding to $V = 300$, closely resembles the pattern in Fig.~\ref{onephase}(a), which shows the one-phase simulation at the same value of surface tension. This similarity is expected, as such a high viscosity ratio effectively mimics one-phase flow. As in Fig.~\ref{onephase}(a), the pattern is highly branched and exhibits substantial finger-length variability. We also verify that the finger bases remain stationary after the initial growth. The saturation profile in Fig.~\ref{twophase}(e) also looks similar to its one-phase counterpart [Fig.~\ref{onephase}(e)], with a clear distinction between the flooded, frozen and active zones.  We measure a fractal dimension of the pattern to be $D = 1.67$, while the best fit of the model yields $\Lambda = 1.5$, with $r_b = 1$ and $E = 0.051$. Note that this viscosity ratio corresponds to the lowest value of $V$ examined by Beeson-Jones \& Woods in their experimental study.

The pattern for $V = 50$ in Fig.~\ref{twophase}(b) retains many of the features observed in the one-phase flow [Fig.~\ref{onephase}(a)] and $V = 300$ [Fig.~\ref{twophase}(a)] cases, but begins to exhibit noticeable deviations. Most prominently, a finger in the north-eastern region shows signs of rupturing and finger self-intersection. Similar effects have been detected in other numerical studies of two-phase flows~\cite{Jackson2015ACell}. This effect appears due to the low value of surface tension considered in the two-phase flow, namely $\sigma = 2\times 10^{-6}$. The overall degree of branching is reduced, and the finger fjords migrate slowly, causing the flooded zone to expand gradually over time. The measured fractal dimension is $D = 1.7$. Interestingly, although the fractal dimension lies within the range reported in the literature, no satisfactory fit, defined as an error $E < 0.1$, was obtained for the saturation curves proposed by the Besson-Jones \& Woods model.

Reducing the viscosity ratio to $V = 25$ leads to substantial finger rupturing and self-intersection, including a case where one finger detaches completely from the main interface. The side-branching seen in Figs.~\ref{twophase}(a) and (b) is absent, and we verified that the finger fjords undergo significant motion during the evolution, rendering the flooded region radius $r_b$ time-dependent. Because the Beeson-Jones-Woods model requires a temporally fixed fully flooded region, it cannot be applied in this regime. The saturation curves confirm that the flooded zone grows in size as time progresses, preventing the formation of a frozen zone. Nevertheless, the saturation still displays an approximately linear dependence on radius at sufficiently large $r$.

For the lowest viscosity ratio considered, $V = 2$, the near-circular interface becomes much less deformed, as expected for such a small ratio. Only shallow indentations appear along the interface, producing very short and broad fingers. The finger bases advance continuously with the interface, as seen in Fig.~\ref{twophase}(h). As a result, no frozen zone forms, and the model cannot be fitted in this regime.

\section{Conclusion}
\label{conclude}

In this work, we employed a level set numerical scheme to assess the validity of the saturation model proposed by Beeson-Jones \& Woods~\cite{Beeson-Jones2019, Beeson-Jones_Thesis}, which aims to describe a universal saturation profile for viscous fingering patterns. Our numerical results show that the Beeson-Jones-Wood model accurately reproduces the saturation profile for one-phase flows when the effective surface tension is sufficiently small and the viscosity ratio sufficiently large. In this regime, the numerically generated patterns exhibit well defined flooded, frozen, and active zones, yielding fitting parameters that are consistent with the experimental data reported in the original study~\cite{Beeson-Jones2019}. This agreement confirms that the model captures the essential dynamics of patterns that closely resemble DLA, when the dimensionless surface tension parameter is small, which corresponds to situations of low surface tension, small plate gaps, large injection rates or a large initial bubble radius.

Alas, the applicability of the model degrades significantly as the effective surface tension increases. We observed that higher surface tension suppresses the frequent branching events and finger length variability necessary to achieve the saturation profile proposed by the Beeson-Jones-Woods model. The less unstable scenario of higher effective surface tensions leads to morphological changes in the resulting patterns, such as ``pinching'' at the finger bases, which may introduce abrupt transitions, local minima, and plateaus in the saturation curves that the Beeson-Jones-Woods model cannot accommodate. The assumption of a time-independent frozen zone breaks for higher surface tensions, as the finger fjords do not remain stationary. From Beeson-Jones' PhD thesis~\cite{Beeson-Jones_Thesis}, there is a suggestion that the saturation model was first developed empirically for DLA simulations and then applied to radial viscous fingering experiments, supporting our conclusion that the model only describes DLA-like scenarios.

Furthermore, our extension of the analysis to two-phase flows demonstrates that the saturation model's validity is restricted to high viscosity ratios, where the system effectively approximates one-phase flow. As the viscosity ratio decreases, the assumption of a stationary flooded region breaks down, as the finger fjords advance radially, and the flooded radius becomes time-dependent. At lower viscosity ratios, phenomena such as finger breakup, coalescence, and the suppression of side branching prevent the formation of a frozen zone entirely. Ultimately, while the Beeson-Jones-Woods model is certainly a valuable tool for characterizing DLA-like interfacial instabilities, it lacks the universality required to describe the full spectrum of nonlinear Hele-Shaw flows, particularly those influenced by significant surface tension or low viscosity contrasts.

\appendix*

\section{Numerical Scheme}
\label{levelset}

For our numerical investigation, we develop an extension to the level set formalism for Hele-Shaw cells proposed by Morrow~\textit{et al.}, which was limited to one-phase flow~\cite{Morrow_Numerical, Morrow_Review, Morrow2023ViscousRotation, Morrow_pof, Morrow_prl}. In the present work, this framework is extended to two-phase flow with a sharp interface.

To numerically solve Eqs.~(\ref{laplacep1})-(\ref{source2}), we construct a level set function $\phi(x,y,t)$ such that the fluid-fluid interface $\partial \Omega(t)$ is the zero level set of $\phi$~\cite{Osher_paper,OsherBook}. In the Hele-Shaw model, as the interfacial evolution is dictated by the velocity in its normal direction, $\phi$ is governed by a level set equation of the form
\begin{equation}
   \label{lsequation}
   \frac{\partial \phi}{\partial t} + F |\nabla \phi| = 0,
\end{equation}
where $F$ is a speed function, continuous and smooth in the entire computational domain, such that $F = v_n$ at $\mathbf{x} \in \partial \Omega(t)$, where $v_n = \mathbf{v}_k \cdot \mathbf{n}$. Due to Eq.~(\ref{kinematic2}), either fluid may be considered when computing $v_n$ at the interface. Except where noted, we consider a uniform square computational grid, of domain $0 \leq r \leq 7.5$ and $0 \leq \theta < 2\pi$.

The level set function $\phi$ is initialized via the method of crossing times, as a distance signed function~\cite{Osher_paper,OsherBook}
\begin{equation}
    \phi = \begin{cases}
        d \; , \; \text{if } \mathbf{x} \in \Omega_2(t), \\
        0 \; , \; \text{if } \mathbf{x} \in \partial \Omega(t), \\
        -d \; , \; \text{if } \mathbf{x} \in \Omega_1(t),
    \end{cases}
\end{equation}
where $d$ is the shortest distance between $\mathbf{x}$ and the interface. This implies that $|\nabla \phi| = 1$. We consider that initially ($t = 0$), the fluid-fluid interface is described by Eq.~(\ref{inital}). Then, $\phi$ is evolved through a second-order total variation diminishing Runge-Kutta scheme, performed by taking two forward Euler steps and then an averaging step, with spatial derivatives being approximated through a second-order essentially nonoscillatory scheme~\cite{Osher_paper,OsherBook}. To maintain numerical stability and accuracy, we choose $\Delta t = \Delta x /(4 \max|F|)$~\cite{Morrow_Review}.

Given the natural numerical error that arises each time step, we periodically perform a re-initialization procedure to ensure $\phi$ is an approximately a distance signed function. Re-initialization is performed every five time steps by solving
\begin{equation}
    \frac{\partial \phi}{\partial \tau} + \text{sgn}(\phi)(|\nabla \phi| - 1) = 0,
\end{equation}
where
\begin{equation}
    \text{sgn}(\phi) = \frac{\phi}{\sqrt{\phi^2 + \Delta x^2}},
\end{equation}
to steady state~\cite{Osher_paper, OsherBook}. Here, $\tau$ is a pseudo time variable where $\Delta \tau = \Delta x/5$.

To evolve Eq.~(\ref{lsequation}), we must compute $F$. The kinematic boundary condition can be written as
\begin{equation}
\label{eqf2}
    F = - \nabla p^{(2)} \cdot \mathbf{n}, \; \mathbf{x} \in \Omega_2(t),
\end{equation}
where $\mathbf{n} = \nabla \phi/|\nabla \phi|$, and $F = v_n$ at the interface, which provides a continuous expression for $F$ in $\Omega_2(t)$. The derivatives in Eq.~(\ref{eqf2}) are evaluated using central differencing. Following Moroney \textit{et. al}~\cite{Moroney2017ExtendingEquation}, we extend $F$ into $\Omega_1$ by solving a biharmonic equation
\begin{equation}
    \nabla^4 F = 0, \; \mathbf{x} \in \Omega_1(t),
\end{equation}
ensuring that $F$ is continuous and differentiable in the entire computational domain, while preserving that $F = v_n$ at $\mathbf{x} \in \partial\Omega(t)$.

To evaluate Eq.~(\ref{eqf2}), we begin by computing the pressure field. We consider Eqs.~(\ref{laplacep1})-(\ref{source2}) in polar coordinates, with $p^{(k)} = p^{(k)}(r,\theta,t)$, with the location of the interface given by $r = s(\theta,t)$. Thus, Laplace's equation becomes~\cite{Morrow_Numerical,Morrow_Review,Osher_paper,OsherBook}
\begin{equation}
    \frac{1}{r} \frac{\partial}{\partial r} \left(r \frac{\partial p^{(k)}}{\partial r} \right) + \frac{1}{r^2} \frac{\partial^2 p^{(k)}}{\partial \theta^2} = \Sigma.
\end{equation}
 For nodes that are not adjacent to the interface, a simple five-point stencil can be applied, such that
\begin{equation}
\label{5stencil}
    \frac{1}{r} \frac{\partial}{\partial r} \left(r \frac{\partial p^{(k)}}{\partial r} \right) \to \frac{1}{r_{i,j} \Delta r} \left( r_{i,j+1/2} \frac{p^{(k)}_{i,j+1} - p^{(k)}_{i,j}}{\Delta r} - r_{i,j-1/2} \frac{p^{(k)}_{i,j} - p^{(k)}_{i,j-1}}{\Delta r} \right),
\end{equation}
where $r_{i,j \pm 1/2} = (r_{i,j\pm1} + r_{i,j})/2 $. The derivatives in the $\theta$ direction are discretised in a similar fashion.

When solving for nodes adjacent to the interface, due to the discontinuity in the pressure, we are unable to apply the stencil of Eq.~(\ref{5stencil}). Suppose that the interface is located at $r = r_I$, where $r_{i,j} < r_I < r_{i,j+1}$, where $r_{i,j} \in \Omega_1$ and $r_{i,j+1} \in \Omega_2$. We define a ghost node~\cite{gibou_2002,gibou2013,chenLS1997} at $r_I$, where the pressure of fluid 1 and 2 are given by $p_I^{(1)}$ and $p_I^{(2)}$, respectively. We can relate these two pressures by rewriting the kinematic boundary condition (\ref{kinematic2}) as
\begin{equation}
\label{kinematicP}
    \nabla p^{(2)} \cdot \mathbf{n} = \frac{1}{V} \nabla p^{(1)} \cdot \mathbf{n}, \; \mathbf{x} \in \partial \Omega(t),
\end{equation}
where $\mathbf{n} = (n_r, n_\theta)$. From Eq.~(\ref{kinematicP}), we assume that
\begin{equation}
\label{kinematicPr}
        \frac{\partial p^{(2)}}{\partial r} = \frac{1}{V} \frac{\partial p^{(1)}}{\partial r}, \; \mathbf{x} \in \partial \Omega(t).
\end{equation}
\begin{equation}
        \frac{\partial p^{(2)}}{\partial \theta} = \frac{1}{V} \frac{\partial p^{(1)}}{\partial \theta}, \; \mathbf{x} \in \partial \Omega(t).
\end{equation}
This procedure was proposed by Osher and Sethian~\cite{Osher_paper,OsherBook}, and while these equations are not generally true, adding $n_r$ multiplied by the first equation to $n_\theta$ multiplied by the second one leads to the correct boundary jump condition, and numerically converges to the correct solution.

Consider that the interface is located between nodes $(i,j) \in \Omega_1$ and $(i,j+1) \in \Omega_2$. Discretising Eq.~(\ref{kinematicPr}) yields
\begin{equation}
\label{kinematicdiscretized}
    \left(\frac{p^{(2)}_{i,j+1} - p_I^{(2)}}{\Delta r - h_n} \right)= \frac{1}{V} \left(\frac{p^{(1)}_I- p_{i,j}^{(1)}}{h_n} \right),
\end{equation}
where
\begin{equation}
    h_n = \Delta r\left|\frac{\phi_{i,j}}{\phi_{i,j} - \phi_{i,j+1}}\right|.
\end{equation}
is the distance between $r_{i,j}$ and $r_I$. On the other hand, the Young-Laplace's condition (\ref{younglaplace2}) imposes
\begin{equation}
    \label{younglaplacediscretized}
    p_I^{(2)} = p_I^{(1)} - \sigma \kappa.
\end{equation}
Combining Eqs.~(\ref{kinematicdiscretized}) and (\ref{younglaplacediscretized}), and solving for $p_I^{(1)}$ yields
\begin{equation}
\label{p1I}
    p_I^{(1)} = \frac{V (\Delta r - h_n) p^{(1)}_{i,j} + h_n (p^{(2)}_{i,j+1} + \sigma \kappa)}{V(\Delta r - h_n) + h_n}.
\end{equation}
Then, as per Chen~\textit{et al.}~\cite{chenLS1997} and Morrow~\textit{et al.}~\cite{Morrow_Numerical,Morrow_Review}, our finite difference stencil at the node $(i,j)$ becomes
\begin{equation}
\frac{1}{r} \frac{\partial}{\partial r} \left(r \frac{\partial p^{(1)}}{\partial r} \right) \to \frac{2}{r_{i,j} (\Delta r + h_n)} \left( \hat{r}_n \frac{p_I^{(1)} - p^{(1)}_{i,j}}{h_n} - r_{i,j-1/2} \frac{p^{(1)}_{i,j} - p^{(1)}_{i,j-1}}{\Delta r}
\right),
\end{equation}
where $\hat{r}_n = (r_{i,j} + r_I)/2$, and $p^{(1)}_I$ is given by Eq.~(\ref{p1I}). Similarly, for the node $(i,j+1)$,
\begin{equation}
    \frac{1}{r} \frac{\partial}{\partial r} \left(r \frac{\partial p^{(2)}}{\partial r} \right) \to \frac{2}{r_{i,j+1}(\Delta r + h_s)}\left(r_{i,j+3/2} \frac{p^{(2)}_{i,j+2} - p^{(2)}_{i,j+1}}{\Delta r} - \hat{r}_s \frac{p^{(2)}_{i,j+1} - p^{(2)}_I}{h_s} \right),
\end{equation}
where $\hat{r}_s = (r_{i,j+1} + r_I)/2$, $h_s = \Delta r|\phi_{i,j+1}/(\phi_{i,j+1}-\phi_{i,j})|$, and
\begin{equation}
    p_I^{(2)} = \frac{V h_s(p^{(1)}_{i,j} - \sigma \kappa) + (\Delta r - h_s) p^{(2)}_{i,j+1}}{V h_s + (\Delta r - h_s)}.
\end{equation}
Note that if we take either one-phase limit, $V = 0$ or $V = \infty$, the discretisation stencil derived by Morrow~\textit{et al.}~\cite{Morrow_Review} is recovered.

If the node and interface are sufficiently close ($h_n < \Delta r^2$), we impose $p^{(k)}_{i,j} = p^{(k)}_I$. An analogous procedure has to be performed if the interface lies between nodes $(i,j-1)$ and $(i,j)$ and for the derivatives in the azimuthal direction. We stress that, given the coupling between the pressure fields in both regions, both $p^{(1)}$ and $p^{(2)}$ must be solved in the entire computational domain at the same time. The generalized curvature is computed as $\kappa = \nabla \cdot \mathbf{n}$ over the entire computational domain, through second-order accurate finite differences. The implementation of the far-field boundary conditions follows the same steps as Morrow~\textit{et. al}~\cite{Morrow_Review, Morrow_Numerical}. For further details on the implementation of the method, we refer the reader to Refs.~\cite{Morrow_Numerical,Morrow_Review, Osher_paper, OsherBook,gibou_2002,gibou2013,chenLS1997}.

\begin{figure}[!ht]
	\centering
	\includegraphics[width = 0.7\textwidth]{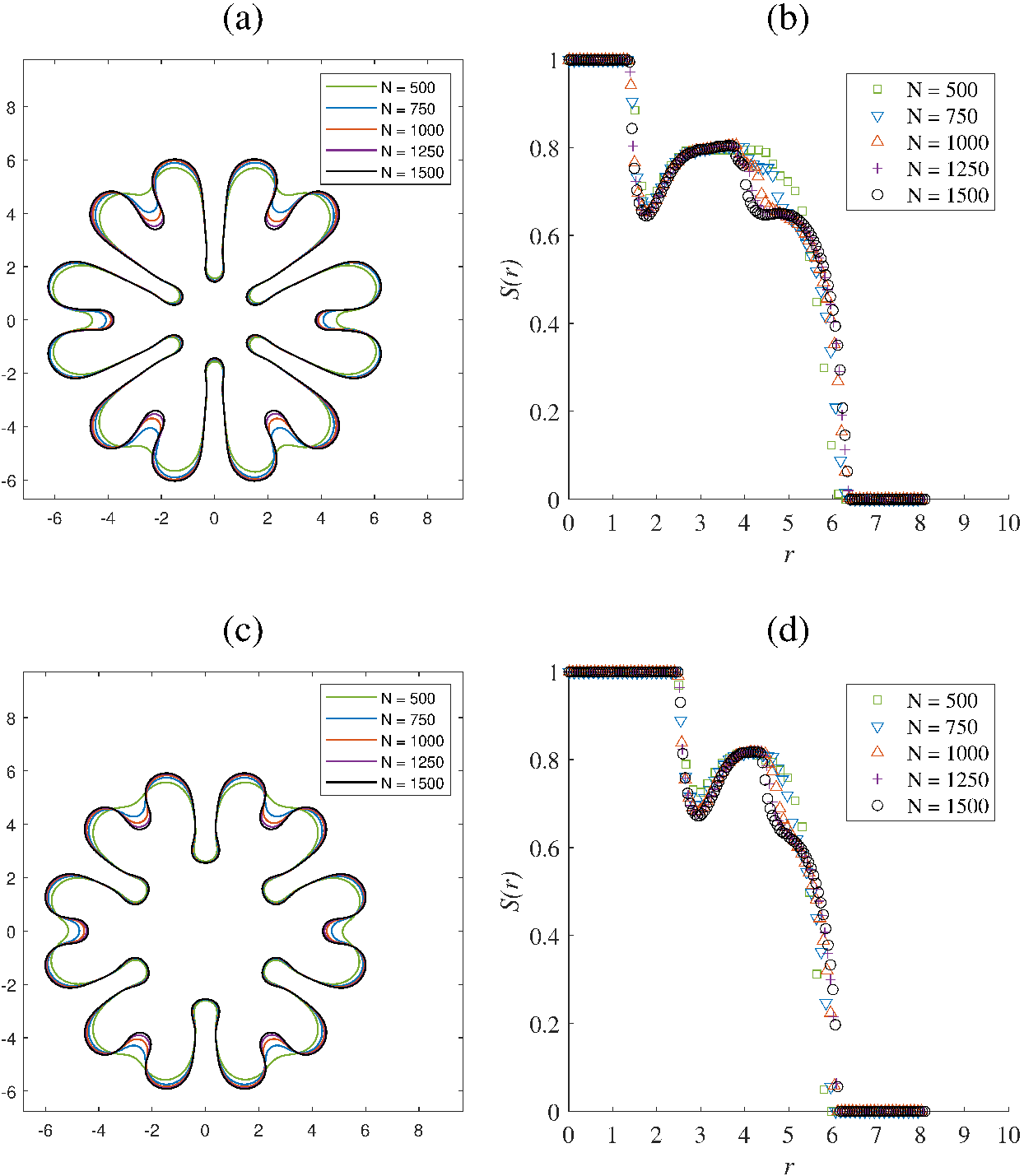}
    	\caption{Snapshots of numerical simulations obtained from the level set method for $\sigma = 1/2000$ at $t = 80$ for $N = 500$, 750, 1000, 1250, and 1500 considering (a) One-phase ($V = \infty$ and (c) two-phase ($V = 10$) flow. The corresponding saturation profiles are presented in (b) for $V = \infty$ and (d) for $V = 10$.}
	\label{convergence}
\end{figure}

Below, we present a convergence test for our scheme. Following a standard convergence test present in several numerical works~\cite{Jackson2015ACell}, in Fig.~\ref{convergence} we present snapshots of numerical simulations for a six-fold symmetric pattern for several grid resolutions, namely $N = 500$, 750, 1000, 1250, and 1500, with an effective surface tension $\sigma = 1/2000$ at $t = 80$. Along with the interfacial patterns in Figs.~\ref{convergence}(a) for one-phase flow ($V = \infty$) and (c) for two-phase flow ($V = 10$), the corresponding saturation profiles $S(r,t)$ are presented in Figs.~\ref{convergence}(b) and (d).

Inspecting Fig.~\ref{convergence}, we see that the point in which the fingers bifurcate is affected by the grid resolution $N$. For smaller $r$ the saturation profile is mostly unaffected by the grid resolution $N$, presenting the formation of a local minima at around $r \approx 1.5$ ($r \approx 2.5$) for $V = \infty$ ($V = 10$). Considering the interfacial patterns, the shape of the saturation curves and the high computational cost of a more refined grid, we chose to consider $N = 1000$ for the entirety of this work. Some simulations were performed for $N = 1250$ and $N = 1500$ and they present the same qualitative behavior as discussed in Figs.~\ref{onephase}-\ref{twophase}.

\begin{acknowledgments}
I.M.C.~wishes to thank the Coordenação de Aperfeiçoamento de Pessoal de N\'ivel Superior - Brasil (CAPES) - Finance Code 001 for financial support through Grant No. 88887.937774/2024-00, and Queensland University of Technology for the invitation as a visitor. S.W.M.~acknowledges support from the Australian Research Council via the Discovery Project DP250101095 and is grateful to Prof.~Andy Woods for introducing this topic at the MATRIX research program ``Instabilities of flow in porous media". We thank Prof.~Jos\'e A.~Miranda for important discussions and useful suggestions.
\end{acknowledgments}

\section*{Data Availability}
The data that support the findings of this article are openly
available \cite{data_scott}

%bibliographystyle{phys}
\bibliography{references}

\end{document}